\documentstyle[12pt]{article}
\begin{document}

\centerline{{\Large\bf Neutrino Mixing and Oscillations}} 

\vspace{.1in}

\centerline{{\Large\bf in Quantum Field Theory  }}

\vspace{.3in}
\centerline{{\bf  E.Alfinito$^{*}$\footnote{alfinito@le.infn.it} , 
M.Blasone$^{\dagger}$\footnote{blasone@vaxsa.csied.unisa.it}, 
A.Iorio$^{\dagger}$\footnote{iorio@vaxsa.csied.unisa.it}
and G.Vitiello$^{\dagger}$\footnote{vitiello@vaxsa.csied.unisa.it \\ \\
Invited talk at the XIXth International Conference on Particle Physics 
and Astrophysics \\
in the Standard Model and Beyond, Bystra, Poland, 
September 19-26, 1995.}}}

\vspace{.2in}
\centerline{${}^{*}$Dipartimento di Fisica dell'Universit\`a}               
\centerline{Via per Arnesano, Lecce} 
\centerline{I-73100 Lecce, Italy}

\vspace{.2in}
\centerline{${}^{\dagger}$Dipartimento di Fisica dell'Universit\`a}         
      \centerline{and INFN, Gruppo Collegato, Salerno} \centerline{I-84100 
Salerno, Italy}

\begin{abstract}
We show that the generator of field mixing transformations in Quantum Field
Theory
induces a non
trivial structure in the vacuum which turns out to be a coherent
state, both for bosons and for fermions, although with a different
condensate structure. 
The Fock space for mixed fields
is unitarily inequivalent to the Fock space of the
massive (free) fields in the infinite volume limit. As a practical 
application we study
neutrino mixing and oscillations.
A new
oscillation formula is found
where the oscillation amplitude is depressed, with respect to the usual 
one, by a factor which is momentum and mass dependent. In the
relativistic limit, the usual formula is recovered. We finally
discuss
in some detail phenomenological features of the modified oscillation 
formula.
\end{abstract}

P.A.C.S. 11.10.-z ,11.30.Jw , 14.60.Gh

\newpage
Novel features of field mixing transformations in Quantum Field
Theory (QFT) have been recently [1,2,3] discovered. 
In particular it has
been shown [1,3] that the generator of such transformations induces a 
non
trivial structure in the vacuum which turns out to be a coherent
state, both for bosons and for fermions, although with a different
condensate structure. 
The Fock space for mixed fields
has been explicitely constructed and it has been shown that,
in the infinite volume limit, it is unitarily inequivalent (orthogonal) to 
the Fock space of the
corresponding massive (free) fields.

As explained below, such new and almost unexpected features find their 
origin in the existence, in QFT, of infinitely many inequivalent
representations of the canonical (anti-)commutation relations [4,5].

The question arises, however, if such a new and rich
structure leads to any new and possibly testable effect. For such a 
purpose,
neutrino mixing and oscillations [6,7] have been investigated in 
ref.[1,2] as a practical example and
a new
oscillation formula (different from the usual one) has been found.
In particular, we have found a correction on the oscillation amplitude 
which turns out to be momentum and mass dependent. However, in the
relativistic limit, the usual formula is recovered; this is in general
agreement with other studies of neutrino oscillations in the non
relativistic region [8].

The aim of the present paper is to report on such results and to discuss
in some detail phenomenological features of the modified neutrino 
oscillation formula.

In the simple case of two flavor mixing [7] ( for the case of three
flavors see ref. 1) the mixing relations are:
$$\nu_{e}(x,t) = \nu_{1}(x,t) \; \cos\theta   + \nu_{2}(x,t) \; \sin\theta  $$
$$\nu_{\mu}(x,t) =- \nu_{1}(x,t) \; \sin\theta   + \nu_{2}(x,t)\; 
\cos\theta\;,  \eqno(1)$$
where $x$ denotes the (three) spatial coordinates; $\nu_{e}(x,t)$ and 
$\nu_{\mu}(x,t)$ are the (Dirac) neutrino fields with definite flavors.
 $\nu_{1}(x,t)$ and $\nu_{2}(x,t)$ are the (free) 
neutrino fields with definite masses $m_{1}$ and $m_{2}$, respectively.
Here we do not need to distinguish between left-handed and right-handed
components.
The fields $\nu_{1}(x,t)$ and $\nu_{2}(x,t)$ are written as 
$$\nu_{i}(x,t) = \frac{1}{\sqrt{V}} \sum_{ k,r}[u^{r}_{k,i}(t)
\alpha ^{r}_{k,i}\:e^{i k x}+ v^{r}_{k,i}(t)
\beta ^{r\dag }_{k,i}\: e^{-i k x}], \; ~ i=1,2 \;. \eqno(2) $$
$\alpha ^{r}_{k,i}$ and $ \beta ^{r }_{k,i}$, $  i=1,2 \;, 
\;r=1,2$ are the annihilator operators for the vacuum state 
$|0\rangle_{1,2}$:
$\alpha ^{r}_{k,i}|0\rangle_{12}= \beta ^{r }_{k,i}|0\rangle_{12}=0$.
For simplicity, we use the same symbol for the vector $k$ and for its 
modulus.
The anticommutation relations are:
$$\{\nu^{\alpha}_{i}(x,t), \nu^{\beta\dag }_{j}(y,t')\}_{t=t'} = 
\delta^{3}(x-y)
\delta _{\alpha\beta} \delta_{ij} \;, \;\;\;\;\; \alpha,\beta=1,..,4 \;,
\eqno(3)$$
and
$$\{\alpha ^{r}_{k,i}, \alpha ^{s\dag }_{q,j}\} = \delta
_{kq}\delta _{rs}\delta _{ij}   ;\qquad \{\beta ^{r}_{k,i}, \beta ^{s\dag
}_{q,j}\} = \delta _{kq} \delta _{rs}\delta _{ij},\;\;\;\; 
i,j=1,2\;. \eqno(4)  $$
All other anticommutators are zero. The orthonormality and 
completeness relations are the usual ones.

Eqs.(1) relate the hamiltonians $H_{1,2}$ (we consider only the mass 
terms) and $H_{e,\mu}$ [7]:
$$H_{1,2}=m_{1}\;\bar {\nu}_{1} \nu_{1} + m_{2}\;\bar {\nu}_{2} \nu_{2} 
\eqno(5)$$  
$$H_{e,\mu}=m_{ee}\; \bar {\nu}_{e} \nu_{e} + 
m_{\mu\mu}\;\bar {\nu}_{\mu} \nu_{\mu}+
m_{e\mu}\left(\bar {\nu}_{e} \nu_{\mu} + \bar {\nu}_{\mu} \nu_{e}\right) 
\eqno(6)$$  
where $m_{ee}=m_{1}\cos^{2}\theta + m_{2} \sin^{2} \theta$, 
$m_{\mu\mu}=m_{1}\sin^{2}\theta + m_{2} \cos^{2} \theta$ 
and $m_{e\mu}=(m_{2}-m_{1})\sin\theta \cos \theta$.

In the LSZ formalism of QFT [4] observables are expressed in terms of 
asymptotic in- (or out-) fields. These fields, also called free or physical 
fields, are obtained by the weak limit of the 
Heisenberg or interacting fields for $t \rightarrow - (or +)
\infty$. The system Lagrangian and the resulting field equations are given 
in terms of the Heisenberg fields and therefore the meaning of the weak 
limit is to provide a realization of the basic dynamics in terms of the 
asymptotic fields. The weak limit is however not unique since 
infinitely many representations of the canonical (anti-)commutation 
relations exist in QFT [4,5] and as a consequence the realization of 
the basic dynamics in terms of the asymptotic fields is not unique. 
Therefore, in order to avoid ambiguities, it is of 
crucial importance to investigate with much care the mapping among 
Heisenberg fields and free fields (generally known as dynamical
mapping or Haag expansion) [4,5].

For example, since unitarily inequivalent representations describe 
physically different phases, in theories with spontaneous symmetry
breaking the same set of Heisenberg field equations describes the
normal (symmetric) phase as well as the symmetry broken phase, according 
to the representation one chooses for the asymptotic fields.

It should be observed that no problem arises
with uniqueness of the asymptotic limit in quantum mechanics, namely for 
finite volume systems. In such a case indeed the von Neumann theorem 
ensures that the representations of the canonical commutation relations 
are each other unitary equivalent.
However, the von Neumann
theorem does not hold in QFT since infinite number of degrees of freedom is 
there considered and much attention is then required when considering any 
mapping among interacting and free fields [4,5].

For these reasons, intrinsic to the QFT structure, the mixing relations 
(1), which can be seen as
a mapping
among Heisenberg fields and free fields,
deserve a
careful analysis.

To this aim, we can rewrite the mixing relations (1) in the form:
$$\nu_{e}^{\alpha}(x,t) = G^{-1}(\theta,t)\; \nu_{1}^{\alpha}(x,t)\; G(\theta,t) $$
$$\nu_{\mu}^{\alpha}(x,t) = G^{-1}(\theta,t)\; \nu_{2}^{\alpha}(x,t)\; G(\theta,t)
\;, \eqno(7)  $$
and the generator $G(\theta,t)$ can be written as: 
$$G(\theta , t) = exp[\theta(S_{+}(t) - S_{-}(t))]\;, \eqno(8)$$
with 
$$ S_{+}(t) \equiv  \int d^{3}x \; \nu_{1}^{\dag}(x,t)
\nu_{2}(x,t) \;\;,\;\;\;
S_{-}(t) \equiv  \int d^{3}x \; \nu_{2}^{\dag}(x,t)
\nu_{1}(x,t)\;= \left( S_{+}(t)\right)^{\dag}\;. \eqno(9) $$
In the following we will omit for
simplicity the time dependence.
It is easy to see, by introducing $ S_{3} 
\equiv \frac{1}{2} \int d^{3}x \left(\nu_{1}^{\dag}(x)\nu_{1}(x) - 
\nu_{2}^{\dag}(x)\nu_{2}(x)\right) $, that the $su(2)$ algebra is closed:
$$ [S_{+} , S_{-}]=2S_{3} \;\;\;,\;\;\; [S_{3} , S_{\pm} ] = \pm S_{\pm}
. \eqno(10)$$

The main point (see ref.[1] for details) is that the above generator of
mixing transformations does not leave invariant the vacuum of the free
fields $\nu_{1,2}$, say $|0 \rangle_{1,2}$, since it induces an $SU(2)$ 
coherent state structure of neutrino-antineutrino pairs in this state
[9,1].
This coherent state is the vacuum for the fields $\nu_{e,\mu}$, which we
denote by $|0 \rangle_{e,\mu}$:
$$|0 \rangle_{e,\mu} = G^{-1}(\theta)\; |0 \rangle_{1,2}\;. \eqno(11)$$

It is then possible [1] to construct explicitely the Fock space for
the mixed operators which can be rewritten in the form:
$$\nu_{e}(x,t) = \frac{1}{\sqrt{V}} \sum_{k,r}\:e^{i k x}
 [u^{r}_{k,1}(t)
\alpha ^{r}_{k,e}(t)\: + v^{r}_{-k,1}(t)
\beta ^{r\dag }_{-k,e}(t)] \eqno(12a)$$
$$\nu_{\mu}(x,t) = \frac{1}{\sqrt{V}} \sum_{k,r}\:e^{i k x}
 [u^{r}_{k,2}(t)
\alpha ^{r}_{k,\mu}(t)\: + v^{r}_{-k,2}(t)
\beta ^{r\dag }_{-k,\mu}(t)] \eqno(12b)$$
where the wave functions for the massive fields have been used [1,3] 
and (in the reference frame  $k=(0,0,|k|)$) the creation and
annihilation operators for the mixed fields are given by:
$$\alpha^{r}_{k,e}(t)=\cos\theta\;\alpha^{r}_{k,1}\;+\;\sin\theta\;\left(
U_{k}^{*}(t)\; \alpha^{r}_{k,2}\;+\;\epsilon^{r}\;
V_{k}(t)\; \beta^{r\dag}_{-k,2}\right) \eqno(13a)$$
$$\alpha^{r}_{k,\mu}(t)=\cos\theta\;\alpha^{r}_{k,2}\;-\;\sin\theta\;\left(
U_{k}(t)\; \alpha^{r}_{k,1}\;-\;\epsilon^{r}\;
V_{k}(t)\; \beta^{r\dag}_{-k,1}\right)  \eqno(13b)$$
$$\beta^{r}_{-k,e}(t)=\cos\theta\;\beta^{r}_{-k,1}\;+\;\sin\theta\;\left(
U_{k}^{*}(t)\; \beta^{r}_{-k,2}\;-\;\epsilon^{r}\;
V_{k}(t)\; \alpha^{r\dag}_{k,2}\right) \eqno(13c)$$
$$\beta^{r}_{-k,\mu}(t)=\cos\theta\;\beta^{r}_{-k,2}\;-\;\sin\theta\;\left(
U_{k}(t)\; \beta^{r}_{-k,1}\;+\;\epsilon^{r}\;
V_{k}(t)\; \alpha^{r\dag}_{k,1}\right) \eqno(13d)$$
with $\epsilon^{r}=(-1)^{r}$ and
$$V_{k}(t)=|V_{k}|\;e^{i(\omega_{k,2}+\omega_{k,1})t}\;\;\;\;,\;\;\;\;
U_{k}(t)=|U_{k}|\;e^{i(\omega_{k,2}-\omega_{k,1})t} \eqno(14)$$
$$|U_{k}|=\left(\frac{\omega_{k,1}+m_{1}}{2\omega_{k,1}}\right)^{\frac{1}{2}}
\left(\frac{\omega_{k,2}+m_{2}}{2\omega_{k,2}}\right)^{\frac{1}{2}}
\left(1+\frac{k^{2}}{(\omega_{k,1}+m_{1})(\omega_{k,2}+m_{2})}\right) 
\eqno(15a)$$
$$|V_{k}|=\left(\frac{\omega_{k,1}+m_{1}}{2\omega_{k,1}}\right)^{\frac{1}{2}}
\left(\frac{\omega_{k,2}+m_{2}}{2\omega_{k,2}}\right)^{\frac{1}{2}}
\left(\frac{k}{(\omega_{k,2}+m_{2})}-\frac{k}{(\omega_{k,1}+m_{1})}\right) 
\eqno(15b)$$
$$|U_{k}|^{2}+|V_{k}|^{2}=1 \eqno(16)$$
$$|V_{k}|^{2}=|V(k,m_{1}, m_{2})|^{2}=
\frac{k^{2}\left[(\omega_{k,2}+m_{2})-(\omega_{k,1}+m_{1})\right]^{2}}
{4\;\omega_{k,1}\omega_{k,2}(\omega_{k,1}+m_{1})(\omega_{k,2}+m_{2})}\eqno
(17)$$
where $\omega_{k,i}=\sqrt{k^{2}+m_{i}^{2}}$.

By using eqs.(13) the expectation value of the number
operator $N_{\sigma_{l}}^{k,r}$ is obtained as:
$$\;_{1,2}\langle0|\;N_{\sigma_{l}}^{k,r}\;|0\rangle_{1,2}\;=
\;\sin^{2}\theta\;|V_{k}|^{2} \;,\;\;\;\; \sigma=\alpha, \beta \;,\;\;\;\;
l=e,\mu ,\eqno(18)$$
Eq.(18) gives the condensation density of the vacuum
state $|0\rangle_{1,2}$
as a function of the mixing angle $\theta$, of the
masses $m_{1}$ and $m_{2}$ and of the momentum modulus $k$, and
it is in contrast with the usual approximation case where one puts
$|0\rangle_{e,\mu}=|0\rangle_{1,2}
\equiv|0\rangle$ and it is $\langle0|\;N_{\alpha_{e}}^{k,r}\;
|0\rangle\;=\;\langle0|\;N_{\alpha_{\mu}}^{k,r}\;
|0\rangle\;=0\;$.
Also note that 
$_{1,2}\langle0|\;N_{\sigma_{l}}^{k,r}\;|0\rangle_{1,2}$
plays the role of zero point contribution when considering
the energy contribution of
${\sigma_{l}}^{k,r}$ particles [1].

The oscillation formula is obtained by using the mixing mappings (13) 
[1]:
$$\langle\alpha_{k,e}^{r}(t)|\;N_{\alpha_{e}}^{k,r}\;
|\alpha_{k,e}^{r}(t)\rangle\;=$$
$$=\;1 -\;\sin^{2}\theta\;|V_{k}|^{2}\;
-\;|U_{k}|^{2}\;\sin^{2}2\theta
\;\sin^{2}\left(\frac{\Delta\omega_{k}}{2}t\right)\;. \eqno(19)$$ 

The fraction of $\alpha_{\mu}^{k,r}$ particles in the same 
state is
$$\langle\alpha_{k,e}^{r}(t)|\;N_{\alpha_{\mu}}^{k,r}\;
|\alpha_{k,e}^{r}(t)\rangle\;=$$
$$=\;|U_{k}|^{2}\;\sin^{2}2\theta
\;\sin^{2}\left(\frac{\Delta\omega_{k}}{2}t\right)+
\;\sin^{2}\theta\;|V_{k}|^{2}\;
\left(1\;-\;\sin^{2}\theta\;|V_{k}|^{2}\right)\;.\eqno(20)$$ 

The terms with $|V_{k}|^{2}$ and $|U_{k}|^{2}$ in (19) and (20) denote the 
contribution from the vacuum condensate.
We have
$$\langle\alpha_{k,e}^{r}(t)|\;N_{\alpha_{e}}^{k,r}\;
|\alpha_{k,e}^{r}(t)\rangle + 
\langle\alpha_{k,e}^{r}(t)|\;N_{\alpha_{\mu}}^{k,r}
\;|\alpha_{k,e}^{r}(t)\rangle=
$$
$$
\langle\alpha_{k,e}^{r}|\;N_{\alpha_{e}}^{k,r}\;
|\alpha_{k,e}^{r}\rangle + 
\langle\alpha_{k,e}^{r}|\;N_{\alpha_{\mu}}^{k,r}
\;|\alpha_{k,e}^{r}\rangle\;. \eqno(21)$$
where $|\alpha_{k,e}^{r}\rangle = |\alpha_{k,e}^{r}(t = 0)\rangle$, which
shows the conservation of the number $(N_{\alpha_{e}}^{k,r}\;
+ N_{\alpha_{\mu}}^{k,r})$ . The expectation value of this number in the 
state $|0\rangle_{1,2}$ is not zero due to the condensate contribution.

Eqs.(19) and (20) are to
be compared with the approximated ones in the conventional treatment:
$$\langle\alpha_{k,e}^{r}(t)|\;N_{\alpha_{e}}^{k,r}\;
|\alpha_{k,e}^{r}(t)\rangle\;=\;
1-\sin^{2}2\theta\;\sin^{2}\left(\frac{\Delta\omega_{k}}{2}t
\right)\;  \eqno(22)$$
and
$$\langle\alpha_{k,e}^{r}(t)|\;N_{\alpha_{\mu}}^{k,r}\;
|\alpha_{k,e}^{r}(t)\rangle\;=\;
\sin^{2}2\theta\;\sin^{2}\left(\frac{\Delta\omega_{k}}{2}t
\right)~, \eqno(23)$$
respectively.

Eqs.(19) and (20) reproduce the conventional ones (22) and (23)
when $|U_{k}|\rightarrow 1$ (and $|V_{k}|\rightarrow 0$).

In conclusion, in the proper QFT treatment we obtain 
corrections to the flavor oscillations which come from the condensate 
contributions. The conventional (approximate) results (22) and (23) are 
recovered when the condensate contributions are missing (in the $|V_{k}| 
\rightarrow 0$ limit).

The phenomenological implications of  
the results (19) and (20) have been discussed in ref.[2] where we have 
studied the function $|V_{k}|^{2}$.

Here we note that $|V_{k}|^{2}$  depends on $k$ only through its modulus 
and it is always in the interval $[0,\frac{1}{2}[$. It has a maximum for 
$k= \sqrt{m_{1}m_{2}}$.
Also, $|V_{k}|^{2} \rightarrow 0$ when $k \rightarrow \infty$.
Moreover, $|V_{k}|^{2}=0$  when $m_{1} = m_{2}$~
(no mixing occurs in Pontecorvo theory in this case).

This last feature is remarkable since the corrections to the oscillations 
depend on the modulus $k$ through $|V_{k}|^{2}$ (and $|U_{k}|^{2} =1-
|V_{k}|^{2}$). So, these corrections disappear in the infinite momentum or 
relativistic
limit $k >> \sqrt{m_{1} m_{2}}$ (note that $\sqrt{m_{1} m_{2}}$ is the
scale of the condensation density). 

However, 
for finite $k$, the oscillation amplitude is depressed, or "squeezed", 
by a factor 
$|U_{k}|^{2}$: the squeezing factor ranges 
from $1$ to $\frac{1}{2}$ depending on $k$ and on the masses values. 
The values of the squeezing factor may therefore
have not
negligible effects in experimental findings and the dependence 
of the flavor oscillation amplitude on the momentum could thus be tested.

To better estimate the effects of the momentum dependence we 
rewrite the $|V_{k}|^{2}$ function as
$$|V_{k}|^{2} \equiv |V(p,a)|^{2}=
\frac{1}{2}\left(1-\frac{1}{\sqrt{1+a\left(\frac{p}{p^{2}+1}\right)^{2}}}
\right)  \eqno(24)$$
with
$$p=\frac{k}{\sqrt{m_{1} m_{2}}}\;\;\;\;\;,\;\;\;\;a=\frac{(\Delta m)^{2}}
{m_{1} m_{2}}\;\;\;,\;\;\;0\leq a < + \infty ~, \eqno(25)$$
where $\Delta m \equiv m_{2}-m_{1}$ (we take $m_{1}\leq m_{2}$).

At $p=1$, $|V(p,a)|^{2}$ reaches its maximum value $|V(1,a)|^{2}$,
which goes 
asymptotically to 1/2 when $a \rightarrow \infty$.

It is useful to calculate the value of $p$, say $ p_{\epsilon}$, at which 
the function $|V(p,a)|^{2}$ becomes a 
fraction $\epsilon$ of its maximum value
$V(1,a)$:
$$p_{\epsilon}=\sqrt{-c+\sqrt{c^{2}-1}}\;\;\;\;,\;\;\;\;
c\equiv\frac{b^{2}(a+2)-2}{2(b^{2}-1)} \;\;\;\;,\;\;\;\;
b\equiv1-\epsilon\left(1-\frac{2}{\sqrt{a+4}}\right)~.\eqno(26)$$

The values of  $\sqrt{ m_{1} m_{2} } $ and of $a$ 
corresponding to some given values of $m_{1}$ and $m_{2}$ chosen below 
the current experimental bounds are reported in Tab. I.

Three sets of values of $|U(p_{\epsilon},a)|^{2}$ and of $k_{\epsilon}$, 
for $\epsilon=1\, , \, \frac{1}{2}\, , \, \frac{1}{10}$, corresponding to 
the values of $m_{1}$ and $m_{2}$ given in Tab. I, are
reported in Tab. II (see also Fig. 1). We used $|U(p_{\epsilon},a)|^{2}=1-
\epsilon+\epsilon \;|U(1,a)|^{2}$ and
$k_{\epsilon}=p_{\epsilon}\; \sqrt{m_{1} m_{2}}$.

We note that neutrinos of not very large momentum may have sensible 
squeezing factors for the oscillation amplitudes.
Larger deviations from the usual oscillation formula may thus be expected
in these low momentum ranges. We note that observations of 
neutrino oscillations by large passive detectors include
neutrino momentum as
low as few hundreds of KeV [6].

We observe that an indication on the
neutrino masses may be given by the dependence,
if experimentally tested,  of the oscillating
amplitude on the 
momentum since the function $|U_{k}|^{2}$ (cf. eqs.(16) and (19)) has a 
minimum at $k= \sqrt{m_{1}m_{2}}$.

Another interesting case not considered in ref.[2] occurs when one of the 
two masses, say
$m_{1}$, goes to zero. In this case, the maximum of the condensation
density (the function $|V_{k}|^{2}$) occurs at $k=0$; however, since 
$a \rightarrow \infty$ when $m_{1} \rightarrow 0$, it is still possible
to have non neglegible effects at rather ``large'' momenta; $m_{2}$ should 
be large in order to provide appreciable corrections. The
situation is illustrated in Tab. III, where for the calculation we used 
$m_{1}=10^{- 10} eV$.

Let us also observe that since the vacuum condensate induces
the correction factor, the vacuum acts as a "momentum (or spectrum) 
analyzer" for the oscillating neutrinos: neutrinos with 
$k\gg\sqrt{m_{1}m_{2}}$ have oscillation amplitude larger
than neutrinos with $k\simeq\sqrt{m_{1}m_{2}}$, 
due to the vacuum structure. Such a 
vacuum spectral analysis effect may sum up to other effects (such as MSW 
effect [10] in the matter) in depressing or enhancing neutrino 
oscillations; in this connection see ref.[1], where the above scheme is also 
generalized to the oscillations in the matter.

On the basis of the above discussion and results we can conclude 
that probing the non relativistic momentum domain seems promising
in order to obtain new insights in neutrino physics.

Further studies on neutrino oscillations in the framework here discussed 
are in progress [11].

Finally, let us mention that the study of the mixing of boson fields 
shows [3] that relations analogous to eqs. (13) and (18) hold 
and the vacuum also acquires a non trivial condensate structure. In the 
boson case we find $|U_{k}| = cosh {\sigma}_{k}$ and $|V_{k}|
= sinh {\sigma}_{k}$ with ${\sigma}_{k} = {1\over{2}}log({{\omega}_{k,1}
\over{\omega}_{k,2}})$ where ${\omega}_{k,i} , i = 1,2$ is the boson 
energy.

We are glad to acknowledge R.Ma\'nka and J.S{\l}adkowski for the
invitation to 
report about our work at the International Conference on Particle Physics 
and Astrophysics in the Standard Model and Beyond, Bystra, September 19-26 
1995, and for their kind hospitality.

\newpage

\begin{center}
Table I:
The values of $\sqrt{ m_{1} m_{2} } 
$ and of $a$ for given values of $m_{1}$ and $m_{2}$.
\end{center}  
\begin{center}
\begin{tabular}{|c||c|c|c|c|} \hline\hline
&{\em $m_{1}$(eV)} & {\em $m_{2}$(KeV)} & {\em $\sqrt{m_{1}m_{2}}$(KeV)} &
{\em $a$}  \\ \hline
$A$ & $ 5   $   &  $ 250 $    &  $   1.12    $    &  $ \sim 5 \cdot 10^{4}  $ 
\\ \hline
$B$ & $ 2.5   $   &  $ 250 $    &  $   0.79    $    &  $ \sim 1 \cdot 10^{5}  $
\\ \hline
$C$ & $ 5   $   &  $ 200 $    &  $   1    $    &  $ \sim 4 \cdot 10^{4}  $ 
\\ \hline
$D$ & $  1  $   &  $ 100 $    &  $  0.32  $  &  $ \sim 1 \cdot 10^{5}  $  \\
\hline
$E$ & $ 0.5 $   &  $ 50  $    &  $  0.15  $  &  $ \sim 1 \cdot 10^{5}  $  \\
\hline
$F$ & $ 0.5 $   &  $ 1   $    &  $  0.02  $  &  $ \sim 2 \cdot 10^{3}  $  \\
\hline
\hline
\end{tabular}
\end{center}

\bigskip
\medskip

\begin{center}
Table II: $|U(p_{\epsilon},a)|^{2} \; $vs.$ \;k_{\epsilon}$.
\end{center} 
\begin{center}
\begin{tabular}{|c||c|c||c|c||c|c|} \hline\hline
 
& {\em $|U(1,a)|^{2}$}           &{\em $k_{1}$(KeV)} 
& {\em $|U(p_{1/2},a)|^{2}$}     & {\em $k_{1/2}$(KeV)}  
& {\em $|U(p_{1/10},a)|^{2}$}    & {\em $k_{1/10}$(KeV)} \\ \hline
$A$ & $ \simeq 0.5 $  &  $ 1.12 $    &  $ \simeq 0.75 $  &  $ \simeq 146 $
&  $ \simeq 0.95 $  &  $ \simeq 519 $ \\ \hline
$B$ & $ \simeq 0.5 $  &  $ 0.79 $    &  $ \simeq 0.75 $  &  $ \simeq 145 $
&  $ \simeq 0.95 $  &  $ \simeq 518 $ \\ \hline
$C$ & $ \simeq 0.5 $  &  $ 1 $    &  $ \simeq 0.75 $  &  $ \simeq 117 $
&  $ \simeq 0.95 $  &  $ \simeq 415 $ \\ \hline
$D$ & $ \simeq 0.5 $  &  $ 0.32 $  &  $ \simeq 0.75 $  &  $ \simeq  58 $
&  $ \simeq 0.95 $  &  $ \simeq 206 $ \\ \hline
$E$ & $ \simeq 0.5 $  &  $ 0.16 $  &  $ \simeq 0.75 $  &  $ \simeq  29 $
&  $ \simeq 0.95 $  &  $ \simeq 104 $ \\ \hline
$F$ & $ \simeq 0.5 $  &  $ 0.02 $  &  $ \simeq 0.75 $  &  $ \simeq 0.6 $
&  $ \simeq 0.95 $  &  $ \simeq  2  $ \\ \hline
\hline
\end{tabular}
\end{center}

\bigskip
\medskip

\begin{center}
Table III: $|U(p_{\epsilon},a)|^{2} \;$ vs.$ \;k_{\epsilon}$ for
$m_{1}\simeq 0$ and different values of $m_{2}$.
\end{center} 
\begin{center}
\begin{tabular}{|c|c||c|c||c|c|} \hline\hline
{\em $m_{1}$(eV) }              &{\em $m_{2}$(KeV)} 
& {\em $|U(p_{1/2},a)|^{2}$}     & {\em $k_{1/2}$(KeV)}  
& {\em $|U(p_{1/10},a)|^{2}$}    & {\em $k_{1/10}$(KeV)} \\ \hline
$ \simeq 0 $  &  $ 250 $  &  $ \simeq 0.75 $  &  $ \simeq 144 $
&  $ \simeq 0.95 $  &  $ \simeq 516 $ \\ \hline
$ \simeq 0 $  &  $ 200 $  &  $ \simeq 0.75 $  &  $ \simeq 115  $
&  $ \simeq 0.95 $  &  $ \simeq 413 $ \\ \hline
$ \simeq 0 $  &  $ 100 $   &  $ \simeq 0.75 $  &  $ \simeq 57 $
&  $ \simeq 0.95 $  &  $ \simeq 206 $ \\ \hline
$ \simeq 0 $  &  $ 50 $  &  $ \simeq 0.75 $  &  $ \simeq 29  $
&  $ \simeq 0.95 $  &  $ \simeq 103 $ \\ \hline
\hline
\end{tabular}
\end{center}

\newpage

{\bf References}

\medskip

\begin{itemize}

\item[1]
 M. Blasone and G. Vitiello, Quantum Field Theory of Fermion 
           Mixing, \\
	SADF1-1995, hep-ph/9501263, {\em Annals of Physics }(N.Y.), 
      in print.

\item[2]
 E.Alfinito, M.Blasone, A.Iorio and G.Vitiello, 
{\it Phys. Lett.}{\bf B 362} 91 (1995).

\item[3]
       M.Blasone and G.Vitiello, {\it Mixing Transformations in Quantum Field 
       	Theory}, in preparation, 1995.

\item[4]
 C. Itzykson and J.B. Zuber, {\it Quantum Field Theory},
	(McGraw-Hill, New York, 1980); \\
N.N. Bogoliubov. A.A. Logunov, A.I. Osak and I.T. Todorov, 
	{\it General Principles of Quantum Field Theory}, 
	(Kluwer Academic Publishers, Dordrech, 1990).

\item[5]
	H. Umezawa, H. Matsumoto and M. Tachiki, 
	{\it Thermo Field Dynamics and Condensed States}, 
(North-Holland Publ.Co., Amsterdam, 1982); \\
	H.Umezawa, {\it Advanced Field Theory: Macro, Micro, and Thermal 
Physics}, 
	(American Institute of Physics, New York, 1993).

\item[6]
R. Mohapatra and P. Pal, {\it Massive Neutrinos in Physics and 
	Astrophysics}, (World Scientific, Singapore, 1991); \\
        J.N. Bahcall, {\it Neutrino Astrophysics}, 
	(Cambridge Univ. Press, Cambridge, 1989). \\
        For an early reference on field mixing see Z. Maki, M. Nakagawa and
	S. Sakata, {\it Prog. Theor. Phys.}{\bf 28} 870 (1962).

\item[7]
S.M. Bilenky and B. Pontecorvo, {\sl Phys. Rep.} {\bf 41} 225 (1978).

\item[8]
   C. Giunti, C.W. Kim and U.W. Lee, {\it Phys. Rev.}{\bf D44} 3635 (1991);\\
        C. Giunti, C.W. Kim, J.A. Lee and U.W. Lee, {\it Phys. Rev.}{\bf D48} 
        4310 (1993);\\
        J. Rich, {\it Phys. Rev.}{\bf D48} 4318 (1993);\\ 
  C. Giunti, C.W. Kim and U.W. Lee, {\it Phys. Rev.}{\bf D45} 2414 (1992). 

\item[9]
	A. Perelomov, {\it Generalized Coherent States and Their
        Applications}, (Springer-Verlag, Berlin, 1986).

\item[10]
	L. Wolfenstein, {\it Phys. Rev.}{\bf D17} 2369 (1978); \\
	S.P. Mikheyev and A.Y. Smirnov, {\it Nuovo Cimento} 
	{\bf 9C} 17 (1986).

\item[11]
       M.Blasone, P.A.Henning  and G.Vitiello, 
{\it Green's Functions and Neutrino Oscillations in Quantum Field 
       	Theory}, in preparation, 1995.         

\end{itemize}


\newpage
\thispagestyle{empty}
\begin{center}

{\bf Figure caption}

\medskip

\bigskip

\bigskip

\bigskip

\end{center}

Fig. 1: The function $|U(p,a)|^{2}$ for the values of parameters of
Tabs.1,2: A (continuous line), B,D,E (dashed line), C (small-dashed line),
F (dotted line).

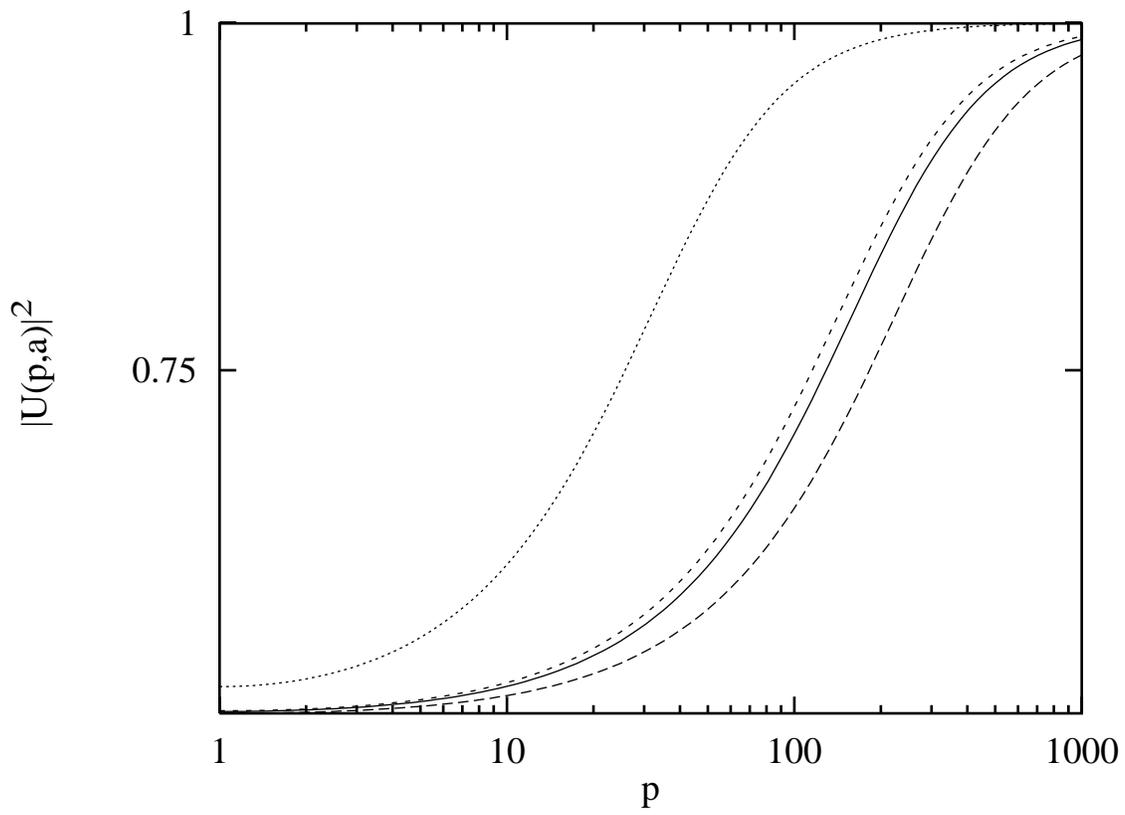
\end{document}